\pdfoutput=1
\DocumentMetadata{} 
\documentclass[sigconf,natbib=true]{acmart}

\author{Yoonseo Kim}
\authornote{These authors contributed equally to this work.}
\email{seo3167@korea.ac.kr}
\affiliation{%
  \institution{Korea University}
  \country{South Korea}
}

\author{Jungwoo Choi}
\authornotemark[1]
\email{wjd030828@g.skku.edu}
\affiliation{%
  \institution{Sungkyunkwan University}
  \country{South Korea}
}

\author{Cheonyoung Park}
\email{park.cheonyoung@kt.com}
\affiliation{%
  \institution{KT Corporation}
  \country{South Korea}
}

\author{Youngwook Kim}
\email{young-wook.kim@kt.com}
\affiliation{%
  \institution{KT Corporation}
  \country{South Korea}
}

\author{Yongho Song}
\email{yongho.song@kt.com}
\affiliation{%
  \institution{KT Corporation}
  \country{South Korea}
}

\author{SeongKu Kang}
\authornote{Corresponding author.}
\email{seongkukang@korea.ac.kr}
\affiliation{%
  \institution{Korea University}
  \country{South Korea}
}

\AtBeginDocument{%
  \providecommand\BibTeX{{%
    \normalfont B\kern-0.5em{\scshape i\kern-0.25em b}\kern-0.8em\TeX}}}
\setcopyright{cc}
\setcctype{by}

\setcopyright{rightsretained}
\copyrightyear{2026}
\acmYear{2026}

\acmConference[KEIR@CIKM 2026]
{The 3rd Workshop on Knowledge-Enhanced Information Retrieval}
{November 8, 2026}
{Rome, Italy}

\acmISBN{}
\acmDOI{}
\acmPrice{}

\acmISBN{}
\acmDOI{}
\acmPrice{}

\usepackage{placeins} 
\usepackage{tcolorbox}       
\usepackage{listings}        
\usepackage{soul}

\usepackage{balance}
\usepackage{pifont}
\usepackage{amsfonts, amsthm, amsmath}
\usepackage[ruled,vlined,linesnumbered]{algorithm2e}
\usepackage{color, colortbl}
\usepackage{enumitem,multirow,graphicx,subcaption,multicol,lipsum,float,adjustbox,makecell}
\newlength{\textfloatsepsave}  
\usepackage{cleveref}
\usepackage{algorithmicx}
\usepackage[page]{appendix} 
 
\crefformat{section}{\S#2#1#3}
\crefformat{subsection}{\S#2#1#3}
\crefformat{subsubsection}{\S#2#1#3}
\usepackage[normalem]{ulem}
\useunder{\uline}{\ul}{}

\usepackage{tabularx}
\usepackage{ragged2e}
\newcolumntype{L}{>{\RaggedRight\arraybackslash}X}

\usepackage{threeparttable}

\newcommand{\proposed}{GRASP\xspace}
\newcommand{\baseline}{SDLBT\xspace}

\newcommand{\smallsection}[1]{{\vspace{0.03in} \noindent \bf {#1}}}

\begin{document}

\title{Concepts Complement Dense Semantics: 
Learning Compact Sparse Spaces for Text-Image Retrieval}

\thanks{Accepted for oral presentation at the 3rd Workshop on Knowledge-Enhanced Information Retrieval (KEIR@CIKM 2026).}

\begin{abstract}
Cross-modal retrieval has been advanced by vision-language pretrained models that encode images and texts into a shared dense embedding space. 
While dense representations effectively capture overall semantic similarity, they often obscure fine-grained visual-textual information needed for precise cross-modal matching.
Recent methods introduce a learned sparse branch to complement dense matching with lexical evidence, but they rely on a redundant language-model token space and lack explicit grounding for sparse dimensions. 
We propose \proposed, a compact and grounded sparse learning framework that mines visual-textual concepts from the corpus. 
A lightweight sparse head is trained to predict concepts relevant to each image or text, yielding interpretable concept-level evidence that complements dense semantic matching.
Extensive experiments show that \proposed improves retrieval accuracy over the state-of-the-art dense-sparse baselines while yielding a more compact and grounded sparse space.
 
\end{abstract}



\maketitle

\section{Introduction}

Cross-modal retrieval is central to multimodal information access, aiming to retrieve relevant images for textual queries or vice versa~\cite{corss_retrieval1,corss_retrieval2,corss_retrieval3,corss_retrieval4,corss_retrieval5}.
Recent progress has been driven by vision-language pre-trained (VLP) models ~\cite{blip, albef}, which encode images and texts into a shared dense embedding space.
By compressing each input as a dense vector, these models effectively capture overall semantic similarity across modalities.
However, dense representations have been reported to underrepresent fine-grained signals, limiting their effectiveness for queries that require precise matching of detailed content, such as specific entities or attributes~\cite{madral, taxoindex, granularity_madral}.




To supply fine-grained signals, prior studies have incorporated explicit knowledge into dense retrievers.
Such knowledge is often derived from domain-specific structures, such as product attributes (e.g., brand, category) in e-commerce search \cite{madral, granularity_madral,reproductive_madral} or academic topics (e.g., fields of study) in scientific search \cite{taxoindex,pairsem}, and combined with dense embeddings to form enriched representations for granular matching.
However, these approaches have largely been limited to single-modal, domain-specific settings.
In cross-modal retrieval, a closely related line of work integrates learned sparse retrieval with dense models \cite{d2s, sdlbt,stair}, where images and texts are additionally projected into the token space of a language model.
By representing inputs over lexical dimensions, the sparse branch captures \textit{token-level} fine-grained information, offering complementary guidance to dense models during both training and retrieval.\footnote{These fine-grained signals can also support retrieval efficiency (e.g., index-based candidate generation~\cite{efficient_inverted_indexes,phrase_inverted_indexes}), but this work focuses on improving final retrieval accuracy.}

%

The state-of-the-art method, Sparse and Dense Learn Better Together (\baseline)~\cite{sdlbt}, encourages dense and sparse branches to learn from each other through bidirectional distillation.
It computes a \textit{teacher} score by combining cross-modal similarities from the dense and sparse branches, and uses this score to supervise both branches.
This joint optimization benefits both branches, particularly improving the standalone retrieval accuracy of the sparse branch.

Still, this design has two limitations.
First, \baseline relies on the full token space of a language model, which is overly redundant for image-text retrieval. 
Many dimensions are weakly related to visual content, including discourse tokens (e.g., ``therefore'', ``however'') and fragmented subword units (e.g., ``\#\#ny'', ``\#\#ism''). 
This bloated space increases computational overhead, obscures meaningful lexical evidence, and degrades interpretability.
We argue that the sparse space should be compact and tailored to retrieval-relevant concepts.

Second, \baseline lacks an explicit mechanism to preserve explicit grounding of the sparse branch, which can cause activated tokens to drift away from meaningful visual-textual concepts.
The sparse projection is largely trained to mimic the teacher score, \textit{without} direct supervision that encourages meaningful lexical tokens to be activated for each image or text.
As a result, the sparse branch can behave like a pseudo-dense representation, improving accuracy through high-dimensional space while losing token-level meaning.
We argue that the sparse branch should be explicitly grounded in meaningful concepts while complementing dense semantics.

\begin{figure*}[th]
    \centering
    \includegraphics[width=0.95\textwidth]{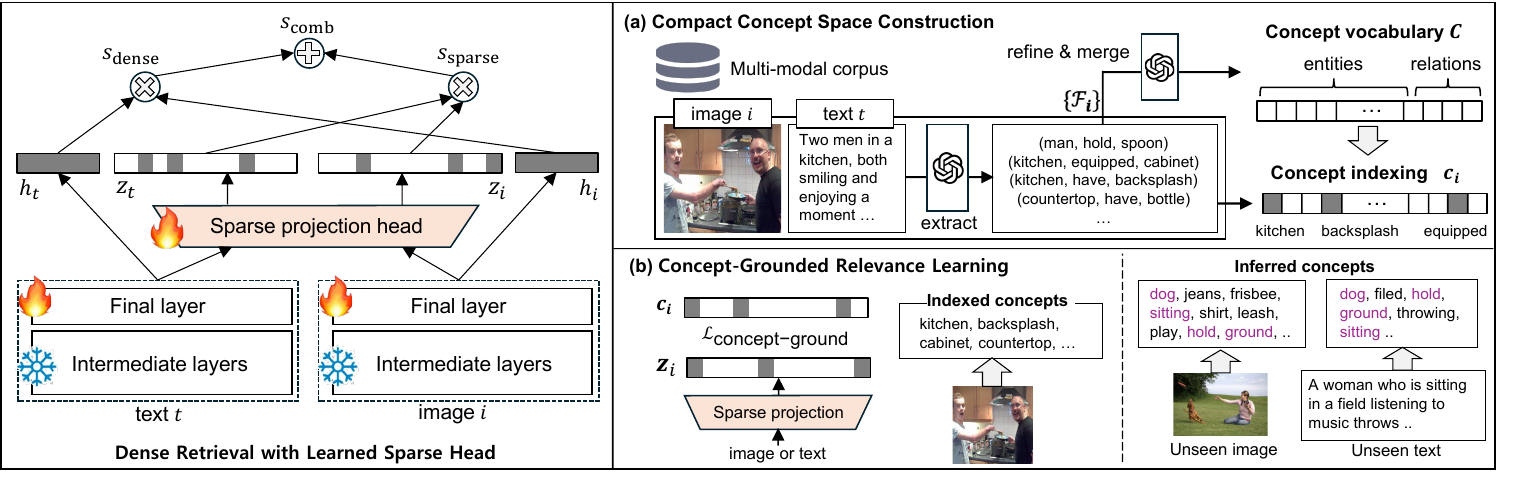}
    \caption{Overview of \proposed framework. We illustrate the image-side process, with the text side handled analogously.}
    \vspace{-0.3cm}
    \label{fig:method}
\end{figure*}

To address these limitations, we propose \textbf{\proposed}, a \textbf{GR}ounded \textbf{A}nd compact \textbf{SP}arse concept framework for text-image retrieval.
Before training, \proposed performs \textit{concept space construction} by mining concepts to bridge visual and textual content from the corpus.
To capture comprehensive contexts, we instantiate these concepts from two complementary aspects: entity- and relation-level units.
The mined concepts define a compact space tailored to cross-modal retrieval, replacing the redundant language-model token space with semantically meaningful sparse dimensions.
During training, \proposed introduces \textit{concept-grounded relevance learning}, where a sparse projection head learns to predict the concepts relevant to each image or text.
This enables \proposed to infer relevant concepts for unseen images or texts, yielding sparse concept representations that complement dense semantic matching while preserving interpretable grounding.
At inference time, \proposed uses the sparse branch as complementary evidence to enhance dense retrieval, adding only negligible cost via the lightweight projection head.

\section{Preliminaries}

\subsection{Problem Formulation}
We consider a cross-modal dataset $\mathcal{D}$ consisting of images and their associated texts (e.g., descriptions). 
For each image $i$, $\mathcal{T}_i$ denotes its associated text set. 
Following~\cite{sdlbt}, we focus on text-to-image retrieval, where a textual query is used to retrieve images in $\mathcal{D}$.
This setting reflects real-world visual content retrieval applications.

\subsection{Dense Retrieval with Learned Sparse Head}
\label{subsec:dense_lexical_distillation}
We describe SDLBT~\cite{sdlbt}, a dense-sparse retrieval framework with a sparse head on top of the backbone VLP, which forms the basis of this work.
Let $f$ denote a VLP model that encodes text and image inputs into contextual embeddings.
Given a text $t$ (or image $i$), $f$ encodes lexical token (or visual patch) sequences into contextual embeddings.
These embeddings are pooled (e.g., CLS or mean pooling) to obtain dense representations $\mathbf{h}_t, \mathbf{h}_i \in \mathbb{R}^d$.

\smallsection{Dense retrieval.}
Dense retrieval computes cross-modal similarity as
$s_{\mathrm{dense}}(t,i) = \mathrm{sim}(\mathbf{h}_t, \mathbf{h}_i)$.
As each input is compressed into a single dense vector, it mainly captures the overall semantic similarity.

\smallsection{Learned sparse branch.}
To supply fine-grained lexical signals, a sparse projection head is added on top of the backbone VLP model.
Specifically, a projection head $g: \mathbb{R}^d \rightarrow \mathbb{R}^{|\mathcal{V}|}$ maps dense representations into a vocabulary-sized token space $\mathcal{V}$.\footnote{Prior work \cite{sdlbt} uses the BERT WordPiece vocabulary ($|\mathcal{V}| \approx 30\mathrm{K}$).}
A sparsifying activation function~\cite{splade1} is then applied to obtain non-negative sparse weights:
$\mathbf{z}_* = \log\left(1 + \mathrm{ReLU}\left(g(\mathbf{h}_*)\right)\right)$.
The sparse similarity is computed as $s_{\mathrm{sparse}}(t,i) = \mathrm{sim}(\mathbf{z}_t, \mathbf{z}_i)$.
Each dimension of $\mathbf{z}_*$ corresponds to a lexical token, allowing the sparse branch to encode token-level matching signals that may be underrepresented in dense vectors.

\smallsection{Integration and optimization.}
Sparse lexical signals complement dense matching through a combined score:
\begin{equation}\label{eq:combine_score}
    s_{\mathrm{comb}}(t,i) = w_{\mathrm{dense}} \cdot s_{\mathrm{dense}}(t,i) + w_{\mathrm{sparse}} \cdot s_{\mathrm{sparse}}(t,i),
\end{equation}
where $w_{\mathrm{dense}}$ and $w_{\mathrm{sparse}}$ control their contributions.
The standard contrastive loss~\cite{infonce} is applied to the dense, sparse, and combined scores.
Further, the combined score is used as a teacher signal to enhance both dense and sparse branches through distillation.
For each $(t,i)$ pair, the learning objective is:
\begin{equation}
    \mathcal{L}_{\mathrm{base}}
    =
    \mathcal{L}_{\mathrm{CL}}
    +
    \mathcal{L}_{\mathrm{distill}},
\end{equation}
where $\mathcal{L}_{\mathrm{CL}}=-\log \frac{\exp(s(t,i))}{\sum_{i' \in \mathcal{B}} \exp(s(t,i'))}$ is applied to all three scores, and $\mathcal{L}_{\mathrm{distill}} = \sum_{s} -p(s_{\mathrm{comb}})\log p(s)$ is applied to dense and sparse scores.
$p(\cdot)$ is the softmax over in-batch candidates.
Regularization and opposite direction (image-to-text) are omitted for simplicity.

\section{METHODOLOGY}
\label{subsec:method}

We follow the dense retrieval with a learned sparse head paradigm.
Our key distinctions lie in two aspects: how we define the sparse space and how we ensure concept grounding of sparse space (Fig.~\ref{fig:method}).



\subsection{Compact Concept Space Construction}
We construct a compact concept space tailored to text-image retrieval. 
The space should bridge the two modalities with fine-grained and interpretable information. 
Inspired by scene graphs~\cite{neural_motifs, Llm4sgg}, which represent a visual scene through entities and their relations, we use entity- and relation-level concepts as language-based units for connecting visual content with textual descriptions.

\smallsection{Concept vocabulary construction.}
Given an image $i$ and its associated text $t \in \mathcal{T}_i$, we instruct an LLM to extract entity-relation-entity triplets, each describing a relation between two entities:
\begin{equation}
    \mathcal{F}_{i,t} = \mathrm{LLM}(i, t\, ;\, P_{\mathrm{extract}}),
    \quad
    \mathcal{F}_i = \textstyle\bigcup_{t \in \mathcal{T}_i} \, \mathcal{F}_{i,t}.
\end{equation}
where $\mathcal{F}_{i,t} = \{(e_1, r, e_2)\}$ is the resulting set for $(i, t)$, and $\mathcal{F}_i$ is the union over all associated texts of image $i$.
$P_{\mathrm{extract}}$ is the prompt.\footnote{The core instruction is: ``Extract visual triplets in the form of (subject, relation, object). Use atomic base nouns for subjects and objects, and use the most direct relation labels''.
When the associated texts are insufficient, $\mathcal{T}_i$ can be augmented by first generating image descriptions and using them as additional textual context (\cref{subsec:exp}).
}


As the extracted information can be inconsistent or noisy, we refine it across the corpus.
We collect all entity and relation phrases and apply $K$-means clustering to their textual embeddings to group semantically similar phrases.
For each cluster $\tilde{\mathcal{C}}_k$, an LLM refines it with prompt
$P_{\mathrm{refine}}$, removing phrases without semantic integrity and
merging only fully substitutable concepts:\footnote{
The core instruction is: ``Remove phrases without semantic integrity or
clear meaning. Merge phrases only when they are fully interchangeable
synonyms''. 
To validate extraction quality, we run RoBERTa-Large-MNLI on Flickr30K captions and their extracted triplets, observing only 1.50\% contradiction, suggesting that the extracted triplets are mostly faithful to the captions.
}
\begin{equation}
\mathcal{C}_k = \mathrm{LLM}(\tilde{\mathcal{C}}_k; P_{\mathrm{refine}}),
\end{equation}

As the dense branch already captures holistic semantics, we define sparse dimensions at the entity/relation level rather than pair/triplet level. 
This keeps the sparse space compact while providing complementary signals.
The final concept vocabulary is defined as $\mathcal{C} = \bigcup_{k=1}^{K} \mathcal{C}_k$, consisting of entities and relations that characterize images in the~corpus.
In our experiments, $|\mathcal{C}|$ is about 4K to 9K, yielding a compact yet expressive set of retrieval-relevant concepts.

\smallsection{Concept indexing.}
Using the concept vocabulary $\mathcal{C}$, we index~concepts (i.e., entities, relations) for both images and texts.
For image $i$, we obtain a concept indicator $\mathbf{c}_i \in \{0,1\}^{|\mathcal{C}|}$ using $\mathcal{F}_i$, where $\mathbf{c}_{ij}=1$ if the $j$-th concept in $\mathcal{C}$ appears in $\mathcal{F}_i$, and $0$ otherwise.
For each text $t \in \mathcal{T}_i$, we obtain $\mathbf{c}_{t}$ in the same way using $\mathcal{F}_{i,t}$.
The resulting indicators contain up to 0.75\% active dimensions in our~experiments.



\subsection{Concept-Grounded Relevance Learning}
\label{subsec:learning_objectives}
Given $\mathcal{C}$, we replace the vast token space of the sparse branch with the concept space, with a projection head $g_{\mathcal{C}}: \mathbb{R}^d \rightarrow \mathbb{R}^{|\mathcal{C}|}$.

\smallsection{Sparse concept branch.}
Unlike \baseline, which projects the pooled representation $\mathbf{h}_*$, \proposed applies the projection \textit{before} pooling to better preserve fine-grained signals. 
Let $\mathbf{h}_{*,m}$ denote the contextual embedding at position $m$, where positions correspond to tokens or patches. 
The sparse concept representation is obtained with the sparsifying activation and max pooling over positions:
\begin{equation}
    \mathbf{z}_{*,j} = \max_{m}
    \log\left(
    1+\mathrm{ReLU}\left(g_{\mathcal{C}}(\mathbf{h}_{*,m})_j\right)
    \right),
    \quad
    j = 1,\ldots,|\mathcal{C}|.
\end{equation}
$g_{\mathcal{C}}(\cdot)_j$ is the raw score for the $j$-th concept.
The max pooling keeps the strongest evidence for each concept across positions, allowing each concept dimension to capture fine-grained evidence from the input.
This yields compact sparse representations $\mathbf{z}_t,\mathbf{z}_i \in \mathbb{R}^{|\mathcal{C}|}$.


\smallsection{Concept grounding objective.}
To ensure that each dimension of $\mathbf{z}_*$ accurately reflects its corresponding concept, we train the sparse branch to predict concepts relevant to each input.
Using the concept-indexed information, we cast this as a multi-label classification task formulated with binary cross-entropy (BCE):
\begin{equation}
\mathcal{L}_{\mathrm{concept-ground}} = \mathcal{L}_\mathrm{BCE}(\mathbf{z}_i, \mathbf{c}_i) + \mathcal{L}_\mathrm{BCE}(\mathbf{z}_t, \mathbf{c}_t).
\end{equation}
This concept grounding loss is applied to both modalities, using each image $i$ and each associated text $t \in \mathcal{T}_i$.

This concept prediction learning provides several benefits. 
First, the sparse branch can infer relevant concepts for unseen images or texts and further supports implicit concept expansion by activating related concepts beyond exact surface matches. 
Moreover, multi-label supervision penalizes irrelevant concept dimensions, leading to sparse and interpretable concept-grounded representations.



\smallsection{Integration and optimization.}
The final retrieval score is obtained by combining the dense and sparse similarities.
For simplicity, we define $s_{\mathrm{comb}}(t,i) = s_{\mathrm{dense}}(t,i) + w \cdot s_{\mathrm{sparse}}(t,i)$, where we set $w=0.1$.
The training process follows the base pipeline in \cref{subsec:dense_lexical_distillation}, except that the sparse branch is guided by the concept grounding loss instead of the contrastive loss:
\begin{equation}
    \mathcal{L}_{\mathrm{\proposed}} = \mathcal{L}_{\mathrm{base}} + \lambda \cdot \mathcal{L}_{\mathrm{concept-ground}},
\end{equation}
where $\lambda$ balances the concept grounding loss. We set $\lambda=0.5$.

\textbf{Remarks on efficiency.}
Concept space construction introduces additional computation, but it is performed only during \textit{offline} indexing phase.
During online retrieval, \proposed adds a lightweight projection head, requiring one matrix multiplication over the stacked contextual embeddings followed by max pooling. 
This introduces negligible overhead compared with the backbone encoder.

\section{Experiments}
\label{subsec:exp}
We follow the setup of the state-of-the-art method, \baseline, including datasets, backbones, and metrics.
As the original image descriptions are often short and limited~\cite{rethink_flickr_coco,eccvcaption}, we augment $\mathcal{T}_i$ with one LLM-generated description per image.
All baselines use the same augmented training data, which is not used for testing.

\smallsection{Datasets and metrics.}
We evaluate on two widely used datasets: Flickr30k~\cite{flickr30k} with 29{,}000/1{,}014/1{,}000 (train / val / test) images and MSCOCO~\cite{coco} with 113{,}287/5{,}000/5{,}000 images, following the Karpathy split~\cite{Karpathy_split}.
The concept vocabulary has $|C|=4{,}046$ on Flickr30k ($\#{\text{ent}}=2{,}964$, $\#{\text{rel}}=1{,}082$) and $|C|=9{,}771$ on MSCOCO ($\#{\text{ent}}=5{,}951$, $\#{\text{rel}}=3{,}820$). 
We report Recall@$\{1, 5\}$ and MRR@10.

\smallsection{Backbone.}
We adopt two VLP backbones, ALBEF~\cite{albef} and BLIP~\cite{blip}, using public checkpoints fine-tuned on each dataset, following~\cite{sdlbt}.


\smallsection{Compared methods.}
We compare \proposed with various recent cross-modal dense-sparse retrieval methods.
(1) \textbf{Dense} denotes a standard dense retriever fully fine-tuned with the augmented data.
(2) \textbf{STAIR}~\cite{stair} maps images and text into a shared sparse token space by grounding visual representations to BERT vocabulary tokens.
(3) \textbf{D2S}~\cite{d2s} maps fixed dense vectors into a sparse token space with probabilistic term expansion.
(4) \textbf{SDLBT}~\cite{sdlbt} is the state-of-the-art method that leverages the combined dense-sparse score as a stronger teacher signal~(\cref{subsec:dense_lexical_distillation}).

We report sparse-only (S), dense-only (D), and combined dense-sparse (D+S) results when applicable.
As \proposed's sparse branch is designed to complement dense retrieval, D+S is our main setting.
For STAIR, whose original model was trained from scratch with a CLIP-style dual encoder, we reimplement its sparse learning method using the same frozen VLP backbones.



\smallsection{Implementation details.}
We train all models with AdamW (batch size 256; patience 20; max 200 epochs), tuning the learning rate over ${10^{-5},10^{-4},10^{-3},10^{-2}}$, with $w=0.1$ and $\lambda=0.5$.
We use \texttt{all-MiniLM-L6-v2} for phrase embeddings, set $K$ to roughly 10 phrases per cluster, and implement $g_{\mathcal{C}}$ as a linear layer with concept-IDF bias initialization.
We use \texttt{InternVL2-26B} for image-description generation and \texttt{GPT-4o mini} for concept construction and LLM-as-a-judge evaluation.
For all baselines, we follow the official implementations and tuning ranges.




\subsection{Results}
\label{sec:Results}


\begin{table}[t] 
\centering
\caption{Performance comparison ($^{\dagger}$: $p<0.05$, paired $t$-test against the best baseline).
LLM-as-judge compares top-1 results against the best baseline following~\cite{chen2024mllm}.}
\label{tab:retrieval_results_detailed}
\renewcommand{\tabcolsep}{0.8mm}
\renewcommand{\arraystretch}{0.95}
\resizebox{\linewidth}{!}{
\begin{tabular}{ll ccc ccc}
\toprule
 & \multirow{2}{*}{\textbf{Method}} & \multicolumn{3}{c}{\textbf{Flickr30k}} & \multicolumn{3}{c}{\textbf{MS-COCO}} \\
\cmidrule(r){3-5} \cmidrule(l){6-8}
& & \textbf{R@1} & \textbf{R@5} & \textbf{MRR@10} & \textbf{R@1} & \textbf{R@5} & \textbf{MRR@10} \\
\midrule
\multirow{9}{*}{\rotatebox{90}{ALBEF}}
& Dense        & 79.65 & 95.49 & 86.58 & 54.65 & 80.46 & 65.67 \\
& STAIR (S)       & 74.98 & 93.86 & 83.05 & 49.76 & 76.92 & 61.34 \\
& STAIR (D+S)     & 79.61 & 95.54 & 86.57 & 54.75 & 80.57 & 65.78 \\
& D2S (S)     & 75.94 & 94.22 & 83.85 & 51.31 & 78.32 & 62.81 \\
& D2S (D+S)     & 79.70 & 95.30 & 86.44 & 53.44 & 79.44 & 64.53 \\
& SDLBT (S)       & 78.61 & 95.27 & 85.79 & 52.97 & 79.20 & 64.14 \\
& SDLBT (D)       & 79.61 & 95.34 & 86.54 & 54.76 & 80.44 & 65.69 \\
& SDLBT (D+S)     & 78.61 & 95.27 & 85.81 & 52.98 & 79.21 & 64.15 \\
& \cellcolor[HTML]{D9D9D9}\proposed & \cellcolor[HTML]{D9D9D9} \textbf{80.81}$^{\dagger}$ & \cellcolor[HTML]{D9D9D9} \textbf{95.92}$^{\dagger}$ & \cellcolor[HTML]{D9D9D9} \textbf{87.38}$^{\dagger}$ & \cellcolor[HTML]{D9D9D9} \textbf{55.47}$^{\dagger}$ & \cellcolor[HTML]{D9D9D9} \textbf{80.70} & \cellcolor[HTML]{D9D9D9} \textbf{66.25}$^{\dagger}$ \\
& \cellcolor[HTML]{D9D9D9} $\,\,$ Sparse only       & \cellcolor[HTML]{D9D9D9} 26.10 & \cellcolor[HTML]{D9D9D9} 50.14 & \cellcolor[HTML]{D9D9D9} 36.35 & \cellcolor[HTML]{D9D9D9} 10.61 & \cellcolor[HTML]{D9D9D9} 26.97 & \cellcolor[HTML]{D9D9D9} 17.61 \\
& \cellcolor[HTML]{D9D9D9} $\,\,$ Dense only        & \cellcolor[HTML]{D9D9D9} 79.75 & \cellcolor[HTML]{D9D9D9} 95.44 & \cellcolor[HTML]{D9D9D9} 86.62 & \cellcolor[HTML]{D9D9D9} 54.49 & \cellcolor[HTML]{D9D9D9} 80.38 & \cellcolor[HTML]{D9D9D9} 65.56 \\
\midrule
\multirow{9}{*}{\rotatebox{90}{BLIP}}
& Dense        & 83.35 & 96.44 & 89.12 & 58.95 & 83.28 & 69.31 \\
& STAIR (S)       & 78.41 & 94.80 & 85.51 & 54.42 & 80.04 & 65.33 \\
& STAIR (D+S)     & 83.34 & 96.52 & 89.10 & 59.02 & \textbf{83.40} & 69.36 \\
& D2S (S)     & 79.66 & 95.55 & 86.47 & 54.73 & 80.62 & 65.67 \\
& D2S (D+S)     & 82.98 & 96.46 & 88.93 & 57.01 & 81.68 & 67.51 \\
& SDLBT (S)       & 81.87 & 96.15 & 88.10 & 57.64 & 82.21 & 68.16 \\
& SDLBT (D)       & 82.91 & 96.49 & 88.83 & 58.62 & 83.09 & 69.02 \\
& SDLBT (D+S)     & 81.89 & 96.17 & 88.10 & 57.67 & 82.21 & 68.16 \\
& \cellcolor[HTML]{D9D9D9}\proposed        & \cellcolor[HTML]{D9D9D9}\textbf{83.95}$^{\dagger}$ & \cellcolor[HTML]{D9D9D9}\textbf{96.92}$^{\dagger}$ & \cellcolor[HTML]{D9D9D9}\textbf{89.61}$^{\dagger}$ & \cellcolor[HTML]{D9D9D9}\textbf{59.40}$^{\dagger}$ & \cellcolor[HTML]{D9D9D9}\textbf{83.40} & \cellcolor[HTML]{D9D9D9}\textbf{69.59}$^{\dagger}$ \\
& \cellcolor[HTML]{D9D9D9}$\,\,$ Sparse only        & \cellcolor[HTML]{D9D9D9}30.79 & \cellcolor[HTML]{D9D9D9}56.26 & \cellcolor[HTML]{D9D9D9}41.71 & \cellcolor[HTML]{D9D9D9}13.63 & \cellcolor[HTML]{D9D9D9}31.80 & \cellcolor[HTML]{D9D9D9}21.46 \\
& \cellcolor[HTML]{D9D9D9}$\,\,$ Dense only        & \cellcolor[HTML]{D9D9D9}83.17 & \cellcolor[HTML]{D9D9D9}96.62 & \cellcolor[HTML]{D9D9D9}89.16 & \cellcolor[HTML]{D9D9D9}58.80 & \cellcolor[HTML]{D9D9D9}83.15 & \cellcolor[HTML]{D9D9D9}69.18 \\
\bottomrule 
\\ 
\toprule
\multicolumn{2}{c}{\multirow{2}{*}{\textbf{LLM-as-judge}}} & \multicolumn{3}{c}{\textbf{Flickr30k}} & \multicolumn{3}{c}{\textbf{MS-COCO}} \\ \cmidrule(r){3-5} \cmidrule(l){6-8}
& & \textbf{Win} & \textbf{Lose} & \textbf{Tie} & \textbf{Win} & \textbf{Lose} & \textbf{Tie} \\ \midrule
\multicolumn{2}{c}{ALBEF}  & $\mathbf{60.0 \pm 0.1\%}$ & $33.0 \pm 0.3\%$ & $7.0 \pm 0.3\%$ & $\mathbf{51.6 \pm 0.6\%}$ & $30.4 \pm 0.2\%$ & $18.0 \pm 0.4\%$ \\
\multicolumn{2}{c}{BLIP}  & $\mathbf{56.8 \pm 0.2\%}$ & $29.3 \pm 0.1\%$ & $13.9 \pm 0.1\%$ & $\mathbf{46.6 \pm 0.1\%}$ & $35.7 \pm 0.9\%$ & $17.8 \pm 0.8\%$ \\
\bottomrule
\end{tabular}}
\end{table}

\subsubsection{\textbf{Overall Evaluation}.}
\Cref{tab:retrieval_results_detailed} reports retrieval performance averaged over three independent runs. 
Overall, \proposed\ consistently achieves the best overall retrieval performance across backbones and datasets, using a substantially smaller concept space than the sparse token-space baselines (i.e., STAIR, D2S, SDLBT).
As noted in prior work \cite{rethink_flickr_coco,eccvcaption}, relevance labels in these benchmarks are often incomplete, leading to \textit{false negatives} where relevant images are not labeled as positives. 
For more comprehensive evaluation, we further conduct an LLM-as-a-judge evaluation~\cite{chen2024mllm}.
\proposed\ consistently achieves a higher winning ratio, outperforming the best baseline by 10.9\% to 27.5\%.

The improvements of \proposed\ come from the complementarity of its sparse branch, not from making it a stronger standalone retriever.
Indeed, its sparse-only performance is relatively low, as expected from the highly compact sparse space and absence of a contrastive loss. Nevertheless, fusion with the dense branch achieves the best performance in every setting.
By contrast, prior sparse branches can show relatively strong sparse-only performance, but they often fail to provide complementary gains after fusion.
For example, SDLBT's BERT-vocabulary sparse branch lowers R@1 below its own dense branch after fusion in all settings.
These results show that the concept-grounded sparse branch captures complementary cross-modal evidence missed by the dense representation.

\begin{figure}[t]
\centering
\includegraphics[width=\linewidth]{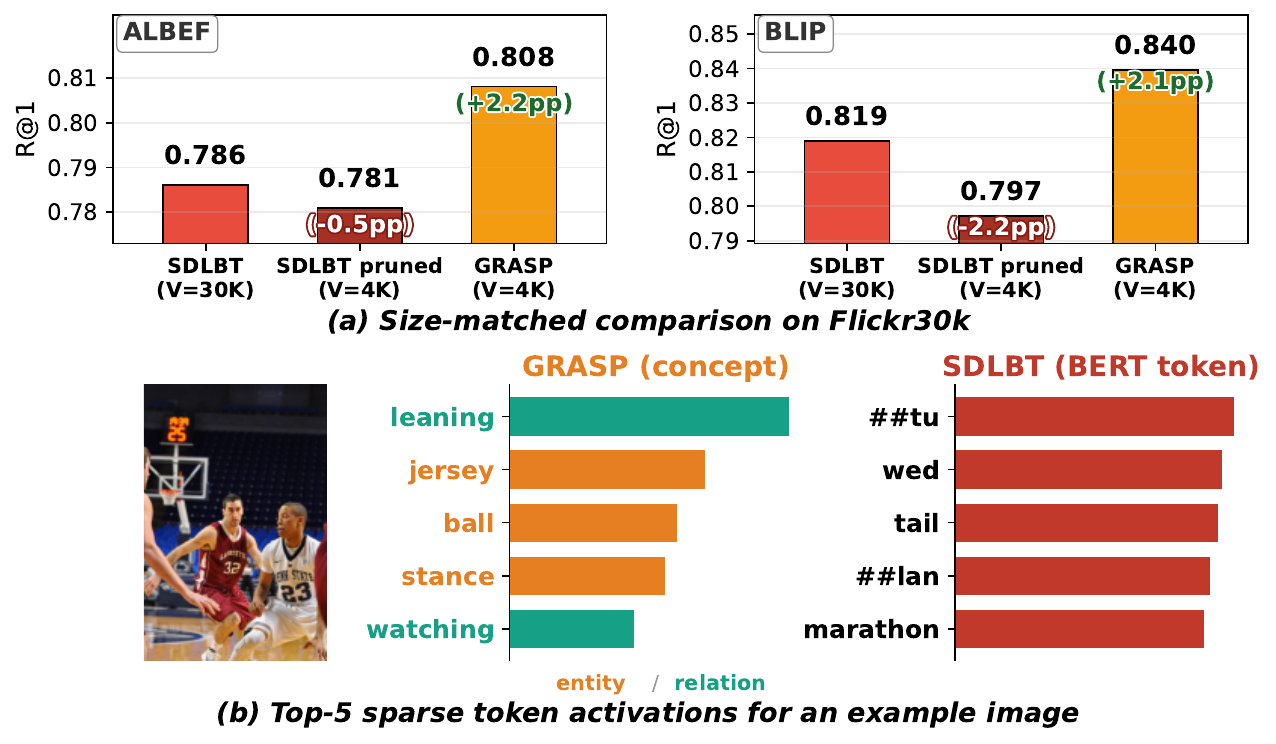}
\caption{In-depth comparison with SDLBT on Flickr30k.}
\label{fig:indepth}
\end{figure}

\begin{table}[t]
\centering
\caption{Design choice analysis on Flickr30k with ALBEF.
The first row corresponds to \baseline.
Active dims denotes the percentage of sparse dimensions activated during inference.
}
\label{tab:ablation_study}
\resizebox{\linewidth}{!}{
\begin{tabular}{lc | cc c}
\toprule
\multirow{2}{*}{\textbf{Concept space}} & \multirow{2}{*}{\shortstack{\textbf{Concept}\\\textbf{grounding}}} & \multirow{2}{*}{\textbf{R@1}}  & \multirow{2}{*}{\shortstack{\textbf{Active dims}\\\textbf{(text/image)}}} & \multirow{2}{*}{\shortstack{\textbf{Sparse}\\\textbf{ dimension}}} \\
 &  &   &  &  \\
\midrule
$\times\,\,$ (BERT tokens)    & $\times$     & 78.61  & 5.8 / 8.1 & 30{,}522 \\
$\checkmark\,\,$ (entity + relation) & $\times$     & 78.91  & 34.3 / 39.9 & 4{,}046  \\
$\checkmark\,\,$ (entity only) & $\checkmark$ & 80.66  & 0.3 / 3.8 & 2{,}964 \\
$\checkmark\,\,$ (entity + relation) & $\checkmark$ & \textbf{80.81} & 0.5 / 4.6 & 4{,}046  \\
\bottomrule
\end{tabular}}
\end{table}

\subsubsection{\textbf{In-depth Comparison with SDLBT}}
\label{sec:Analysis}
We compare \proposed\ and SDLBT in sparse space design and grounding in \Cref{fig:indepth}.

\noindent\textbf{Sparse space design.}
We conduct a size-matched comparison by pruning SDLBT’s BERT token space to the \textit{top-activated 4K tokens}, matching the dimensionality of our concept space.
As shown in \Cref{fig:indepth}(a), this simple pruning substantially degrades SDLBT, while \proposed outperforms both the full 30K SDLBT and the 4K pruned variant.
This validates the effectiveness of our sparse space design.

\noindent\textbf{Sparse activation grounding.}
We analyze the top activated sparse dimensions for an image to examine whether sparse activations are meaningfully grounded.
As discussed earlier, \baseline lacks direct grounding supervision.
Indeed, \baseline activates fragmented or weakly grounded tokens (e.g., \textsf{\#\#tu}, \textsf{\#\#lan}).
In contrast, \proposed\ activates interpretable object and relation concepts relevant to the image (e.g., \textsf{jersey}, \textsf{leaning}), showing that concept grounding provides compact and interpretable evidence.


\subsubsection{\textbf{Design Choice Analysis}}
\Cref{tab:ablation_study} analyzes the contribution of the two main design choices: concept space construction and concept grounding.
First, replacing the 30K BERT token space with our compact 4K concept space lifts R@1 from $78.61$ to $78.91$ while reducing the sparse dimensionality by $7.5\times$.
This suggests that the compact space compresses redundant token dimensions, reducing computational cost while preserving retrieval-relevant signals.

Furthermore, concept grounding improves R@1 to $80.81$, showing that explicitly aligning sparse dimensions with their corresponding concepts is crucial for fully leveraging the compact concept space.
Notably, concept grounding makes the sparse representation substantially more selective: the portion of active dimensions drops from $34.3\%/39.9\%$ to $0.5\%/4.6\%$ for text/image.
This shows that explicit grounding encourages selective activation of concepts relevant to each input.
The entity-only variant ablates relation concepts, and the full entity+relation space achieves the best performance.

\section{Conclusion}
\label{sec:conclusion}
We present \proposed, a text-image retrieval framework that addresses two limitations of prior dense-sparse retrieval: a redundant token space and insufficient grounding of sparse dimensions.
\proposed\ introduces \emph{concept space construction} to replace the large token vocabulary, and \emph{concept-grounded relevance learning} to explicitly anchor sparse dimensions to their target concepts.
Across Flickr30k and MS-COCO with ALBEF and BLIP backbones, \proposed\ consistently outperforms the state-of-the-art method while using a substantially smaller sparse space.
In future work, we plan to extend concept-grounded sparse learning to more diverse retrieval scenarios.

\section*{Acknowledgments}
This work was the result of project supported by KT (Korea Telecom)-Korea University AICT R\&D Center.
This work was also supported by the IITP grant funded by the MSIT (IITP-2026-RS-2026-25616664, AI Star Fellowship Support Program).

\section*{G\lowercase{en}AI U\lowercase{sage} D\lowercase{isclosure}}
\label{sec:genai}
In compliance with ACM's Authorship Policy and CIKM submission guidelines, we disclose that generative AI tools were selectively employed during this research. 
Specifically, \textbf{InternVL2-26B} was utilized to augment insufficient textual contexts by generating image descriptions. 
For concept vocabulary construction, \textbf{GPT-4o mini} was employed to extract entity-relation-entity triplets and refine these concepts within clusters. 
Finally, \textbf{GPT-4o mini} served as an automated judge to evaluate top-1 results via pairwise comparisons. 
No AI-generated text, data, or technical conclusions were incorporated without rigorous human validation, and the authors remain fully accountable for all content and work integrity.

\bibliographystyle{ACM-Reference-Format}
\balance
\bibliography{acmart}

@inproceedings{blip,
title={Blip: Bootstrapping language-image pre-training for unified vision-language understanding and generation},
author={Li, Junnan and Li, Dongxu and Xiong, Caiming and Hoi, Steven},
booktitle={International conference on machine learning},
pages={12888--12900},
year={2022},
organization={PMLR}
}

@article{albef,
  title={Align before fuse: Vision and language representation learning with momentum distillation},
  author={Li, Junnan and Selvaraju, Ramprasaath and Gotmare, Akhilesh and Joty, Shafiq and Xiong, Caiming and Hoi, Steven Chu Hong},
  journal={Advances in neural information processing systems},
  volume={34},
  pages={9694--9705},
  year={2021}
}

@inproceedings{sdlbt,
  title={Sparse and Dense Retrievers Learn Better Together: Joint Sparse-Dense Optimization for Text-Image Retrieval},
  author={Song, Jonghyun and Lee, Youngjune and Cho, Gyu-Hwung and Song, Ilhyeon and Kim, Saehun and Jo, Yohan},
  booktitle={Proceedings of the 34th ACM International Conference on Information and Knowledge Management},
  pages={5268--5272},
  year={2025}
}

@inproceedings{splade1,
  title={Splade: Sparse lexical and expansion model for first stage ranking},
  author={Formal, Thibault and Piwowarski, Benjamin and Clinchant, St{\'e}phane},
  booktitle={Proceedings of the 44th International ACM SIGIR Conference on Research and Development in Information Retrieval},
  pages={2288--2292},
  year={2021}
}

@inproceedings{madral,
  title={Multi-aspect dense retrieval},
  author={Kong, Weize and Khadanga, Swaraj and Li, Cheng and Gupta, Shaleen Kumar and Zhang, Mingyang and Xu, Wensong and Bendersky, Michael},
  booktitle={Proceedings of the 28th ACM SIGKDD Conference on Knowledge Discovery and Data Mining},
  pages={3178--3186},
  year={2022}
}

@inproceedings{granularity_madral,
  title={A multi-granularity-aware aspect learning model for multi-aspect dense retrieval},
  author={Sun, Xiaojie and Bi, Keping and Guo, Jiafeng and Yang, Sihui and Zhang, Qishen and Liu, Zhongyi and Zhang, Guannan and Cheng, Xueqi},
  booktitle={Proceedings of the 17th ACM International Conference on Web Search and Data Mining},
  pages={674--682},
  year={2024}
}

@inproceedings{pairsem,
  title     = {PairSem: LLM-Guided Pairwise Semantic Matching for Scientific Document Retrieval},
  author    = {Kweon, Wonbin and Tian, Runchu and Kang, Seongku and Jiang, Pengcheng and Lu, Zhiyong and Han, Jiawei and Yu, Hwanjo},
  booktitle = {Proceedings of the ACM Web Conference 2026},
  pages     = {2396--2407},
  year      = {2026}
}

@inproceedings{taxoindex,
  title={Taxonomy-guided semantic indexing for academic paper search},
  author={Kang, SeongKu and Zhang, Yunyi and Jiang, Pengcheng and Lee, Dongha and Han, Jiawei and Yu, Hwanjo},
  booktitle={Proceedings of the 2024 Conference on Empirical Methods in Natural Language Processing},
  pages={7169--7184},
  year={2024}
}

@inproceedings{d2s,
  title={Multimodal learned sparse retrieval with probabilistic expansion control},
  author={Nguyen, Thong and Hendriksen, Mariya and Yates, Andrew and Rijke, Maarten de},
  booktitle={European Conference on Information Retrieval},
  pages={448--464},
  year={2024},
  organization={Springer}
}

@article{infonce,
  title={Representation learning with contrastive predictive coding},
  author={Oord, Aaron van den and Li, Yazhe and Vinyals, Oriol},
  journal={arXiv preprint arXiv:1807.03748},
  year={2018}
}

@inproceedings{neural_motifs,
  title={Neural motifs: Scene graph parsing with global context},
  author={Zellers, Rowan and Yatskar, Mark and Thomson, Sam and Choi, Yejin},
  booktitle={Proceedings of the IEEE conference on computer vision and pattern recognition},
  pages={5831--5840},
  year={2018}
}

@inproceedings{Llm4sgg,
  title={Llm4sgg: Large language models for weakly supervised scene graph generation},
  author={Kim, Kibum and Yoon, Kanghoon and Jeon, Jaehyeong and In, Yeonjun and Moon, Jinyoung and Kim, Donghyun and Park, Chanyoung},
  booktitle={Proceedings of the IEEE/CVF Conference on Computer Vision and Pattern Recognition},
  pages={28306--28316},
  year={2024}
}

@inproceedings{coco,
  title={Microsoft coco: Common objects in context},
  author={Lin, Tsung-Yi and Maire, Michael and Belongie, Serge and Hays, James and Perona, Pietro and Ramanan, Deva and Doll{\'a}r, Piotr and Zitnick, C Lawrence},
  booktitle={European conference on computer vision},
  pages={740--755},
  year={2014},
  organization={Springer}
}

@inproceedings{flickr30k,
  title={Flickr30k entities: Collecting region-to-phrase correspondences for richer image-to-sentence models},
  author={Plummer, Bryan A and Wang, Liwei and Cervantes, Chris M and Caicedo, Juan C and Hockenmaier, Julia and Lazebnik, Svetlana},
  booktitle={Proceedings of the IEEE international conference on computer vision},
  pages={2641--2649},
  year={2015}
}

@inproceedings{Karpathy_split,
  title={Deep visual-semantic alignments for generating image descriptions},
  author={Karpathy, Andrej and Fei-Fei, Li},
  booktitle={Proceedings of the IEEE conference on computer vision and pattern recognition},
  pages={3128--3137},
  year={2015}
}

@inproceedings{chen2024mllm,
  title={Mllm-as-a-judge: Assessing multimodal llm-as-a-judge with vision-language benchmark},
  author={Chen, Dongping and Chen, Ruoxi and Zhang, Shilin and Wang, Yaochen and Liu, Yinuo and Zhou, Huichi and Zhang, Qihui and Wan, Yao and Zhou, Pan and Sun, Lichao},
  booktitle={Forty-first International Conference on Machine Learning},
  year={2024}
}

@inproceedings{rethink_flickr_coco,
  title={Rethinking benchmarks for cross-modal image-text retrieval},
  author={Chen, Weijing and Yao, Linli and Jin, Qin},
  booktitle={Proceedings of the 46th international ACM SIGIR conference on research and development in information retrieval},
  pages={1241--1251},
  year={2023}
}

@inproceedings{eccvcaption,
  title={Eccv caption: Correcting false negatives by collecting machine-and-human-verified image-caption associations for ms-coco},
  author={Chun, Sanghyuk and Kim, Wonjae and Park, Song and Chang, Minsuk and Oh, Seong Joon},
  booktitle={European conference on computer vision},
  pages={1--19},
  year={2022},
  organization={Springer}
}

@inproceedings{stair,
  title={Stair: Learning sparse text and image representation in grounded tokens},
  author={Chen, Chen and Zhang, Bowen and Cao, Liangliang and Shen, Jiguang and Gunter, Tom and Jose, Albin and Toshev, Alexander and Zheng, Yantao and Shlens, Jonathon and Pang, Ruoming and others},
  booktitle={Proceedings of the 2023 Conference on Empirical Methods in Natural Language Processing},
  pages={15079--15094},
  year={2023}
}

@inproceedings{reproductive_madral,
  title={Reproducibility Analysis and Enhancements for Multi-aspect Dense Retriever with Aspect Learning},
  author={Bi, Keping and Sun, Xiaojie and Guo, Jiafeng and Cheng, Xueqi},
  booktitle={European Conference on Information Retrieval},
  pages={194--209},
  year={2024},
  organization={Springer}
}

@inproceedings{corss_retrieval1,
  title={Cross-modal implicit relation reasoning and aligning for text-to-image person retrieval},
  author={Jiang, Ding and Ye, Mang},
  booktitle={Proceedings of the IEEE/CVF conference on computer vision and pattern recognition},
  pages={2787--2797},
  year={2023}
}

@inproceedings{corss_retrieval2,
  title={Ei-clip: Entity-aware interventional contrastive learning for e-commerce cross-modal retrieval},
  author={Ma, Haoyu and Zhao, Handong and Lin, Zhe and Kale, Ajinkya and Wang, Zhangyang and Yu, Tong and Gu, Jiuxiang and Choudhary, Sunav and Xie, Xiaohui},
  booktitle={Proceedings of the IEEE/CVF conference on computer vision and pattern recognition},
  pages={18051--18061},
  year={2022}
}

@article{corss_retrieval3,
  title={Cross-modal retrieval: a systematic review of methods and future directions},
  author={Wang, Tianshi and Li, Fengling and Zhu, Lei and Li, Jingjing and Zhang, Zheng and Shen, Heng Tao},
  journal={Proceedings of the IEEE},
  volume={112},
  number={11},
  pages={1716--1754},
  year={2025},
  publisher={IEEE}
}

@article{corss_retrieval4,
  title={USER: Unified semantic enhancement with momentum contrast for image-text retrieval},
  author={Zhang, Yan and Ji, Zhong and Wang, Di and Pang, Yanwei and Li, Xuelong},
  journal={IEEE Transactions on Image Processing},
  volume={33},
  pages={595--609},
  year={2024},
  publisher={IEEE}
}

@inproceedings{corss_retrieval5,
  title={Make: Vision-language pre-training based product retrieval in taobao search},
  author={Zheng, Xiaoyang and Wang, Zilong and Li, Sen and Xu, Ke and Zhuang, Tao and Liu, Qingwen and Zeng, Xiaoyi},
  booktitle={Companion Proceedings of the ACM Web Conference 2023},
  pages={356--360},
  year={2023}
}

@inproceedings{efficient_inverted_indexes,
  title={Efficient inverted indexes for approximate retrieval over learned sparse representations},
  author={Bruch, Sebastian and Nardini, Franco Maria and Rulli, Cosimo and Venturini, Rossano},
  booktitle={Proceedings of the 47th International ACM SIGIR Conference on Research and Development in Information Retrieval},
  pages={152--162},
  year={2024}
}

@inproceedings{phrase_inverted_indexes,
  title={Inverted indexes for phrases and strings},
  author={Patil, Manish and Thankachan, Sharma V and Shah, Rahul and Hon, Wing-Kai and Vitter, Jeffrey Scott and Chandrasekaran, Sabrina},
  booktitle={Proceedings of the 34th international ACM SIGIR conference on Research and development in Information Retrieval},
  pages={555--564},
  year={2011}
}

\end{document}